\documentclass[lettersize,journal]{IEEEtran}
\usepackage{amsmath,amsfonts,amssymb}
\usepackage{algorithmic}
\usepackage{algorithm}
\usepackage{array}
\usepackage[caption=false,font=normalsize,labelfont=sf,textfont=sf]{subfig}
\usepackage{textcomp}
\usepackage{stfloats}
\usepackage{url}
\usepackage{verbatim}
\usepackage{graphicx}
\usepackage{cite}
\begin{document}

\title{Environment-Aware Resource Allocation for Pinching-Antenna-Assisted EDMA-NOMA Systems}

\author{Yaxuan Luo%
\thanks{Y. Luo graduated from the School of Engineering, The University of Manchester, Manchester, U.K., in 2025. Email: nancie7luoyaxuan@gmail.com.}}

\markboth{}%
{Luo: Environment-Aware Resource Allocation for Pinching-Antenna-Assisted EDMA-NOMA Systems}

\IEEEpubid{}

\maketitle

\begin{abstract}
Environment division multiple access (EDMA) exploits line-of-sight (LoS) availability, blockage diversity, and spatial isolation in the propagation environment to regulate inter-region interference, while non-orthogonal multiple access (NOMA) improves intra-region access efficiency through power-domain multiplexing and successive interference cancellation (SIC). This paper investigates heterogeneous environment-map-aware user matching and power allocation for pinching-antenna-assisted EDMA-NOMA systems. Based on large-scale path loss and an exponential LoS blockage model, an average LoS/NLoS effective large-scale channel gain is constructed, allowing the environment map to affect both desired service links and inter-region interference links. This avoids the overly optimistic assumption that environmental blockage only suppresses interference. A utility-based joint user-pairing and power-allocation algorithm (UBA-JPPA) is then proposed to jointly account for system throughput and the scheduled-user Jain fairness index under NOMA power constraints and an SIC-consistent rate definition. Simulation results show that, compared with the ablation baseline without a heterogeneous environment map, the proposed scheme provides consistent gains in sum rate, fairness, and the throughput--fairness operating region. Meanwhile, a strong EDMA-OMA benchmark remains competitive in some high-SNR regimes, indicating that the focus of this work is to verify the benefit of heterogeneous environment information for EDMA-NOMA resource allocation rather than to claim that NOMA universally outperforms OMA.
\end{abstract}

\begin{IEEEkeywords}
EDMA, NOMA, pinching antenna, environment-aware communications, heterogeneous environment map, user matching, power allocation.
\end{IEEEkeywords}

\section{Introduction}

Future sixth-generation (6G) wireless networks are expected to support dense connectivity while simultaneously meeting requirements on high throughput, fairness, and manageable computational complexity. Power-domain NOMA allows multiple users to share the same time-frequency resource block via superposition transmission and separates user signals at the receiver through SIC. It has therefore been widely investigated as a multiple access technique for improving spectral efficiency \cite{saito_noma,benjebbour_noma,ding_noma,noma_survey,islam_noma_survey}. Nevertheless, the performance of NOMA is highly dependent on user pairing and power allocation. When the effective channel disparity between paired users is insufficient, or when the power allocation is inappropriate, both SIC reliability and weak-user fairness may degrade \cite{ding_noma,timotheou_fairness}.

Recently, reconfigurable propagation environments, near-field multiple access, and flexible antenna systems have introduced new multiple-access dimensions beyond the conventional time, frequency, code, space, and power domains \cite{ouyang_ngma,ding_pinching}. Pinching antennas activate controllable radiating points along a dielectric waveguide, providing a new hardware foundation for reconfiguring LoS links, reducing large-scale path loss, and regulating inter-region interference \cite{ding_pinching,wang_noma_pass,zhao_wdma}. In this context, EDMA exploits LoS blockage and spatial isolation in the propagation environment to form relatively independent service regions, thereby suppressing multiuser interference \cite{ding_edma}.

Existing studies on EDMA mainly focus on forming environmentally isolated regions through pinching-antenna position selection, whereas studies on NOMA-assisted pinching-antenna systems typically focus on antenna activation or superposition-transmission performance \cite{ding_pinching,wang_noma_pass}. Therefore, how a heterogeneous environment map can jointly guide EDMA region association, NOMA user pairing, and power allocation remains insufficiently studied. This issue is particularly relevant in indoor corridors, warehouses, shelf environments, and partitioned spaces, where non-uniform blockage distributions imply that distance-based channel gains alone may not accurately reflect link-specific LoS/NLoS propagation differences and inter-region leakage risks.

To address these issues, this paper proposes a heterogeneous environment-map-aware resource allocation framework for EDMA-NOMA systems. Each EDMA region schedules one NOMA user pair, and both service links and inter-region interference links are characterized by environment-map-weighted effective large-scale channel gains. The main contributions of this paper are summarized as follows: i) A downlink pinching-antenna-assisted EDMA-NOMA system model is established, incorporating large-scale path loss, heterogeneous LoS blockage, average LoS/NLoS effective channel gains, and imperfect SIC into the achievable-rate calculation. ii) An environment-map-assisted LoS probability model is introduced, making the local blockage factor a resource-allocation dimension distinct from distance-based path loss; meanwhile, an NLoS residual power coefficient is used to capture the joint impact of blockage on both desired and interfering links. iii) A UBA-JPPA algorithm is designed to jointly consider throughput, scheduled-user fairness, and SIC consistency during user pairing and power search, with local swap search further improving the global utility. iv) Ablation experiments and heterogeneity-sensitivity analysis demonstrate that the proposed scheme provides consistent yet moderate gains over a structurally identical baseline without the environment map, indicating that the performance improvement mainly comes from the explicit use of heterogeneous environment-map information.

\section{System Model}

\subsection{Hybrid EDMA-NOMA System}

Consider a downlink multiuser EDMA-NOMA system in which the base station deploys $M$ pinching antennas along a segmented dielectric waveguide. Each pinching antenna corresponds to one EDMA region. The location of the $m$-th pinching antenna is denoted by
$\mathbf q_m=(x_m^{\rm PA},0,d)$,
and the location of user $k$ is denoted by
$\mathbf u_k=(x_k,y_k,0)$.
The $m$-th EDMA region schedules one NOMA user pair $\mathcal P_m=\{s_m,w_m\}$, where $s_m$ and $w_m$ denote the strong and weak users with respect to region $m$, respectively. The strong--weak user relationship is determined by the effective large-scale channel gain defined later. The original distance-dependent large-scale path gain follows the commonly used distance-attenuation model \cite{rappaport_wireless,3gpp38901,ding_pinching}
\begin{equation}
    g_{k,m}=\left(\frac{d_0}{d_{k,m}}\right)^\alpha,
    \quad
    d_{k,m}=\|\mathbf u_k-\mathbf q_m\|_2,
    \label{eq:pathloss}
\end{equation}
where $d_0$ is the reference distance and $\alpha$ is the path-loss exponent. Equation \eqref{eq:pathloss} is a normalized large-scale power-gain model. Since this work focuses on the impact of environment maps on user matching and power allocation, small-scale fading, waveguide propagation loss, antenna radiation patterns, and dynamic map errors are not explicitly modeled. In practical systems, these factors can be absorbed into equivalent channel gains or considered as extensions for robust optimization.

\subsection{Heterogeneous Environment Map and Effective LoS/NLoS Channel Model}

LoS/NLoS states are commonly characterized by distance-dependent probability models or scenario-dependent empirical models \cite{bai_heath_mmwave,3gpp38901}. To describe non-uniform blockage, this paper defines an environment-map matrix
\begin{equation}
    \mathbf B=[\beta_{k,m}]\in\mathbb R^{K\times M},
    \label{eq:envmap}
\end{equation}
where $\beta_{k,m}$ represents the long-term environmental blockage strength of the link between user $k$ and pinching antenna $m$. Unlike resource allocation that relies only on distance-based path loss, this work treats the environment map $\mathbf B$ as an additional input, extending the resource-allocation decision rule from $\mathcal F(g_{k,m})$ to $\mathcal F(g_{k,m},\beta_{k,m})$. As a result, user matching and power allocation depend on both propagation distance and environmental blockage heterogeneity.

Based on this map, the LoS probability is modeled as
\begin{equation}
    p_{k,m}^{\rm LoS}=\exp\left(-\phi \beta_{k,m} d_{k,m}\right),
    \label{eq:plos}
\end{equation}
where $\phi$ denotes the global blockage intensity. The units of the parameters are selected such that $\phi\beta_{k,m}d_{k,m}$ is dimensionless. This exponential LoS model is consistent with the idea of distance-dependent LoS probability modeling under random blockage \cite{bai_heath_mmwave}. It should be emphasized that \eqref{eq:plos} is not intended to replace the 3GPP scenario model; rather, it augments distance-dependent LoS probability modeling with an environment-map weight to capture the fact that links with the same distance may have different LoS availabilities \cite{bai_heath_mmwave,3gpp38901}.

To isolate the contribution of the environment map, a homogeneous distance-only LoS model is used as an ablation baseline:
\begin{equation}
    p_{k,m}^{\rm Hom}=\exp\left(-\phi d_{k,m}\right).
    \label{eq:phom}
\end{equation}
The blockage coefficient $\beta_{k,m}$ is categorized into three link types, $\beta_{\rm open}$, $\beta_{\rm normal}$, and $\beta_{\rm blocked}$, satisfying
$\beta_{\rm open}<\beta_{\rm normal}<\beta_{\rm blocked}$. A smaller $\beta_{k,m}$ indicates higher LoS availability, such as that in an open corridor, whereas a larger $\beta_{k,m}$ represents stronger blockage in shelf-, wall-, or obstacle-dense regions.

To avoid the overly idealized assumption that environmental blockage only suppresses interference without weakening the desired link, this paper further adopts an average LoS/NLoS effective large-scale gain model. Under the heterogeneous environment map, the effective gain is defined as
\begin{equation}
    \bar g_{k,m}^{\rm Het}
    =g_{k,m}\left[p_{k,m}^{\rm LoS}+\chi\left(1-p_{k,m}^{\rm LoS}\right)\right],
    \label{eq:geffhet}
\end{equation}
where $\chi\in[0,1]$ is the NLoS residual power coefficient. When $\chi=1$, LoS and NLoS links have the same average power, whereas $\chi=0$ corresponds to a completely unavailable NLoS link. In the simulations, $\chi=0.1$, corresponding to approximately $10$ dB NLoS power loss. Accordingly, the homogeneous effective gain without the heterogeneous environment map is
\begin{equation}
    \bar g_{k,m}^{\rm Hom}
    =g_{k,m}\left[p_{k,m}^{\rm Hom}+\chi\left(1-p_{k,m}^{\rm Hom}\right)\right].
    \label{eq:geffhom}
\end{equation}
Unless otherwise specified, rate evaluation uses the true heterogeneous effective gain $\bar g_{k,m}^{\rm Het}$. The ablation scheme without the environment map uses $\bar g_{k,m}^{\rm Hom}$ only during matching and power optimization, while its final performance is still evaluated using the true heterogeneous effective gain. In practical systems, $\beta_{k,m}$ can be obtained from field measurements, radio-map construction, sensing-assisted environment reconstruction, or long-term LoS/NLoS statistics \cite{3gpp38901,bai_heath_mmwave}. If map estimation errors are considered, the estimated blockage coefficient can be written as
\begin{equation}
    \hat{\beta}_{k,m}=\beta_{k,m}(1+\epsilon_{k,m}),
    \label{eq:betaerror}
\end{equation}
where $\epsilon_{k,m}$ is the relative estimation error. Robust optimization under map uncertainty is left for future work.

\subsection{Downlink NOMA Transmission Model}

The transmitted signal in EDMA region $m$ is the power-domain NOMA superposition signal \cite{saito_noma,benjebbour_noma,ding_noma,noma_survey}
\begin{equation}
    x_m=\sqrt{a_{w,m}P_m^{\rm tx}}s_{w,m}
    +\sqrt{a_{s,m}P_m^{\rm tx}}s_{s,m},
    \label{eq:txsignal}
\end{equation}
where $P_m^{\rm tx}$ is the transmit power of region $m$, $s_{w,m}$ and $s_{s,m}$ are unit-power information symbols, $a_{w,m}+a_{s,m}=1$, and downlink NOMA typically allocates no less power to the weak user than to the strong user, i.e., $a_{w,m}\geq a_{s,m}$ \cite{ding_noma,noma_survey}. The normalized transmit SNR is defined as
\begin{equation}
    \rho=\frac{P_m^{\rm tx}}{N_0B},
    \label{eq:rho}
\end{equation}
where $N_0B$ is the noise power. This paper assumes that all EDMA regions use the same transmit power, i.e., $P_m^{\rm tx}=P^{\rm tx}$; hence, a common $\rho$ is used throughout to denote the normalized transmit SNR of each EDMA region. If region-dependent transmit powers are adopted, $\rho$ can be directly generalized to a region-specific $\rho_m$.

For notational simplicity, $\bar g_{k,m}$ denotes the effective large-scale gain used for rate evaluation in this subsection. For the proposed scheme and final performance evaluation, $\bar g_{k,m}=\bar g_{k,m}^{\rm Het}$. The weak user $w_m$ directly decodes its own signal and treats the strong user's signal as intra-pair interference. Its SINR is
\begin{equation}
    \gamma_{w,m}=\frac{a_{w,m}\rho \bar g_{w_m,m}}
    {a_{s,m}\rho \bar g_{w_m,m}+I_{w,m}+1}.
    \label{eq:sinrweak}
\end{equation}
Since the LoS/NLoS environmental effect has been embedded in the effective gain $\bar g_{k,m}$, the inter-region interference term is written as
\begin{equation}
    I_{w,m}=\eta\sum_{n\neq m}\rho \bar g_{w_m,n},
    \label{eq:interweak}
\end{equation}
where $\eta\geq0$ is an inter-region interference scaling factor that captures the leakage strength among different EDMA regions. Equation \eqref{eq:interweak} does not explicitly distinguish the power coefficients of the strong and weak users in region $n$, because the out-of-region interference is determined by the total power of the NOMA superposition signal in that region, and $a_{s,n}+a_{w,n}=1$. Therefore, the average leakage interference from region $n$ to user $w_m$ can be represented by $\rho\bar g_{w_m,n}$.

The strong user first decodes the weak user's message for SIC. The corresponding SINR is \cite{ding_noma,timotheou_fairness}
\begin{equation}
    \gamma_{s\rightarrow w,m}=\frac{a_{w,m}\rho \bar g_{s_m,m}}
    {a_{s,m}\rho \bar g_{s_m,m}+I_{s,m}+1},
    \label{eq:sicweak}
\end{equation}
where
\begin{equation}
    I_{s,m}=\eta\sum_{n\neq m}\rho \bar g_{s_m,n}.
    \label{eq:interstrong}
\end{equation}
After imperfect SIC, the SINR for the strong user to decode its own signal is \cite{noma_survey,islam_noma_survey}
\begin{equation}
    \gamma_{s,m}=\frac{a_{s,m}\rho \bar g_{s_m,m}}
    {\xi a_{w,m}\rho \bar g_{s_m,m}+I_{s,m}+1},
    \label{eq:sinrstrong}
\end{equation}
where $\xi\in[0,1]$ is the residual SIC factor. Here, $\xi=0$ represents ideal SIC, whereas $\xi>0$ means that part of the weak user's signal remains as residual interference.

Based on Shannon-type achievable-rate expressions \cite{shannon1948,cover_thomas}, the direct decoding rate and the SIC decoding rate are respectively given by
\begin{equation}
    R_{u,m}=B\log_2(1+\gamma_{u,m}),\quad u\in\{s,w\},
    \label{eq:userrate}
\end{equation}
\begin{equation}
    R_{s\rightarrow w,m}=B\log_2(1+\gamma_{s\rightarrow w,m}).
    \label{eq:sicrate}
\end{equation}
Here, $R_{w,m}$ denotes the weak user's direct decoding rate and is also denoted by $R_{w,m}^{\rm dir}$ when necessary to avoid confusion with the effective weak-user rate. To ensure that the weak-user codeword can be decoded both by the weak user and by the strong user's SIC receiver, the effective weak-user rate is defined as
\begin{equation}
    R_{w,m}^{\rm eff}=\min\left\{R_{w,m},R_{s\rightarrow w,m}\right\}.
    \label{eq:effweak}
\end{equation}
Equation \eqref{eq:effweak} is a commonly used SIC-consistent rate treatment in downlink NOMA: the weak user's coding rate must not exceed the decoding rate supported by the strong user during SIC \cite{ding_noma,timotheou_fairness}.

The system sum throughput is then
\begin{equation}
    R_{\rm sum}=\sum_{m=1}^{M}\left(R_{w,m}^{\rm eff}+R_{s,m}\right).
    \label{eq:sumrate}
\end{equation}
The Jain fairness index is used to measure the rate balance among scheduled users \cite{jain}:
\begin{equation}
    J_{\rm sch}=\frac{\left(\sum_{k\in\mathcal{S}}R_k\right)^2}
    {|\mathcal{S}|\sum_{k\in\mathcal{S}}R_k^2},
    \label{eq:jain}
\end{equation}
where $\mathcal S$ denotes the $2M$ users scheduled in each Monte Carlo realization, and the weak-user rate is computed using $R_{w,m}^{\rm eff}$. To avoid ambiguity, the Jain fairness index shown in the subsequent figures refers to the scheduled-user fairness $J_{\rm sch}$; the rates of unscheduled candidate users are not included in the main fairness curves. For nonnegative rates, $J_{\rm sch}\in[1/|\mathcal S|,1]$, and a value closer to one indicates a more balanced rate allocation among scheduled users.

\section{Problem Formulation}

This paper aims to jointly optimize user matching and power allocation in the heterogeneous environment-aware EDMA-NOMA system. The optimization variables include the user-pair set $\mathcal P=\{\mathcal P_1,\ldots,\mathcal P_M\}$ and the power-allocation variables $\mathbf a=\{a_{s,m},a_{w,m}\}_{m=1}^{M}$. Unlike conventional NOMA user pairing, the considered problem accounts not only for the service-link effective channel-gain disparity but also for the inter-region leakage differences induced by the heterogeneous propagation environment. Specifically, $\bar g_{k,m}$ affects the intra-region NOMA decoding performance, whereas $\bar g_{k,n}$ determines the average interference experienced by user $k$ from other regions.

To jointly characterize throughput and fairness, this paper adopts a weighted-sum scalarization of the multi-objective utility \cite{boyd_convex}:
\begin{equation}
    U(\mathcal{P},\mathbf a)
    =(1-\lambda)\frac{R_{\rm sum}}{R_{\rm ref}}+\lambda J_{\rm sch},
    \label{eq:globalutility}
\end{equation}
where $\lambda\in[0,1]$ is the throughput--fairness trade-off weight, and $R_{\rm ref}>0$ is a rate-normalization constant that makes the throughput term comparable in scale to $J_{\rm sch}$. The same $R_{\rm ref}$ is used for all schemes, and hence $\lambda$ has a consistent meaning across different schemes. When $\lambda$ approaches zero, the objective emphasizes system throughput; when $\lambda$ approaches one, it emphasizes scheduled-user fairness.

The joint optimization problem is formulated as
\begin{align}
    \max_{\mathcal{P},\mathbf a} \quad
    & (1-\lambda)\frac{R_{\rm sum}}{R_{\rm ref}}+\lambda J_{\rm sch}
    \label{eq:opt}\\
    \text{s.t.}\quad
    & a_{s,m}+a_{w,m}=1,\quad \forall m, \label{eq:c1}\\
    & 0<a_{s,m}\leq a_{w,m}<1,\quad \forall m, \label{eq:c2}\\
    & \mathcal{P}_m\cap\mathcal{P}_n=\emptyset,\quad m\neq n, \label{eq:c3}\\
    & |\mathcal{P}_m|=2,\quad \forall m. \label{eq:pairsize}
\end{align}
Constraints \eqref{eq:c1} and \eqref{eq:c2} impose the normalized power-allocation rule and the higher-power allocation principle for the weak user in downlink power-domain NOMA. Constraint \eqref{eq:c3} ensures that no user is scheduled more than once, and \eqref{eq:pairsize} indicates that each EDMA region schedules one NOMA user pair. SIC consistency is guaranteed by \eqref{eq:effweak} and further encouraged in the algorithm through a soft penalty term. Since the problem contains both combinatorial matching variables and continuous power variables, exhaustive search is computationally demanding. Therefore, a low-complexity approximate algorithm is developed.

\section{Proposed Heterogeneous Environment-Aware UBA-JPPA Algorithm}

This section presents the UBA-JPPA algorithm. Since directly optimizing \eqref{eq:globalutility} requires joint enumeration of user--region matching and power allocation, this paper uses a pair-level utility function as a low-complexity local surrogate for the global objective.

For a candidate user pair $(i,j)$ in region $m$, the strong and weak users are determined by the effective service-link gain:
\begin{equation}
    s=\arg\max_{u\in\{i,j\}}\bar g_{u,m},\quad
    w=\arg\min_{u\in\{i,j\}}\bar g_{u,m}.
    \label{eq:swdefinition}
\end{equation}
This rule follows the basic downlink NOMA design principle that the strong user performs SIC while the weak user receives higher power \cite{ding_noma,noma_survey}. The heterogeneous environment map affects both service-link quality and inter-region interference through the effective gain $\bar g_{k,m}$, thereby influencing SINR, achievable rate, and the final pairing utility.

Given $a_s$ and $a_w=1-a_s$, the rates are computed according to \eqref{eq:sinrweak}--\eqref{eq:effweak}. The candidate-pair sum rate and local Jain fairness index are respectively defined as
\begin{equation}
    R_{i,j}^{m}=R_{s,m}+R_{w,m}^{\rm eff},
    \label{eq:pairrate}
\end{equation}
\begin{equation}
    J_{i,j}^{m}=\frac{\left(R_{s,m}+R_{w,m}^{\rm eff}\right)^2}
    {2\left(R_{s,m}^2+\left(R_{w,m}^{\rm eff}\right)^2\right)}.
    \label{eq:localjain}
\end{equation}
To discourage pairings that are unfavorable for SIC, a decoding-rate-gap-based SIC penalty is introduced. Let $R_{w,m}^{\rm dir}=B\log_2(1+\gamma_{w,m})$ denote the weak user's direct decoding rate. Then,
\begin{equation}
    C_{i,j}^{\rm SIC}
    =\frac{\max\left(0,R_{w,m}^{\rm dir}-R_{s\rightarrow w,m}\right)}{R_{\rm ref}^{m}},
    \label{eq:sicpenalty}
\end{equation}
where $R_{\rm ref}^{m}=R_{\rm ref}/M$. If $R_{s\rightarrow w,m}<R_{w,m}^{\rm dir}$, the strong user cannot support the weak user's direct coding rate during SIC, and the candidate pair is penalized. Since the actual weak-user rate is still limited by \eqref{eq:effweak}, the penalty in \eqref{eq:sicpenalty} only guides the matching process and does not alter rate feasibility.

The local environment-aware utility of candidate triplet $(m,i,j)$ is defined as
\begin{equation}
    U_{m,i,j}
    =\max_{a_s\in\mathcal A}
    \left[(1-\lambda)\frac{R_{i,j}^{m}}{R_{\rm ref}^{m}}
    +\lambda J_{i,j}^{m}
    -\mu C_{i,j}^{\rm SIC}\right],
    \label{eq:localutility}
\end{equation}
where $\mathcal A$ is the strong-user power-search set and $\mu\geq0$ is the SIC penalty weight. The three terms in \eqref{eq:localutility} respectively encourage higher pair throughput, more balanced intra-region rates, and pairings more favorable for SIC. The environment map affects the utility through the following chain:
\begin{equation}
    \beta_{k,n}\rightarrow p_{k,n}^{\rm LoS}\rightarrow \bar g_{k,n}
    \rightarrow I_{k,m}\rightarrow \gamma_{k,m}\rightarrow R_{k,m}\rightarrow U_{m,i,j}.
    \label{eq:envchain}
\end{equation}
Therefore, user matching is not determined solely by distance-based path loss; it is also affected by the heterogeneous LoS/NLoS effective propagation structure.

Algorithm \ref{alg:ubajppa} summarizes the main steps of UBA-JPPA. First, $g_{k,m}$, $p_{k,m}^{\rm LoS}$, and $\bar g_{k,m}^{\rm Het}$ are computed for all user--region links. Then, for the unassigned region and user sets, the algorithm enumerates candidate triplets $(m,i,j)$ and obtains their maximum local utilities through power-grid search. In each round, the region--user pair with the highest utility is scheduled, and the corresponding region and users are removed from the candidate sets. After the greedy assignment, local swap search is performed; a swap is accepted only if it improves the global utility in \eqref{eq:globalutility}. Finally, the power coefficients are refined.

\begin{algorithm}[t]
\caption{Proposed UBA-JPPA Algorithm}
\label{alg:ubajppa}
\begin{algorithmic}[1]
\REQUIRE User set $\mathcal K$, EDMA region set $\mathcal M$, environment map $\{\beta_{k,m}\}$, power-search set $\mathcal A$, weights $\lambda$ and $\mu$.
\ENSURE Scheduled user pairs $\mathcal P$ and power coefficients $\mathbf a$.
\STATE Compute $g_{k,m}$, $p_{k,m}^{\rm LoS}$, and $\bar g_{k,m}^{\rm Het}$ for all $k\in\mathcal K$ and $m\in\mathcal M$.
\STATE Initialize the unassigned user set $\mathcal U\leftarrow\mathcal K$, the unassigned region set $\mathcal R\leftarrow\mathcal M$, and $\mathcal P\leftarrow\emptyset$.
\WHILE{$\mathcal R\neq\emptyset$}
    \FOR{each $m\in\mathcal R$ and each user pair $(i,j)\subset\mathcal U$}
        \STATE Determine $s$ and $w$ according to \eqref{eq:swdefinition}.
        \STATE Search over $a_s\in\mathcal A$ and compute $U_{m,i,j}$ according to \eqref{eq:localutility}.
    \ENDFOR
    \STATE Select $(m^\star,i^\star,j^\star)=\arg\max_{m,i,j}U_{m,i,j}$.
    \STATE Schedule $(i^\star,j^\star)$ in region $m^\star$ and store the corresponding power coefficient.
    \STATE Update $\mathcal U\leftarrow\mathcal U\setminus\{i^\star,j^\star\}$ and $\mathcal R\leftarrow\mathcal R\setminus\{m^\star\}$.
\ENDWHILE
\STATE Perform local swap search among scheduled users; accept a swap only if it improves \eqref{eq:globalutility}.
\STATE Refine the power coefficients of all scheduled user pairs over $\mathcal A$.
\RETURN $\mathcal P$ and $\mathbf a$.
\end{algorithmic}
\end{algorithm}

For the ablation baseline without the environment map, the algorithmic flow remains unchanged, but $\bar g_{k,m}^{\rm Het}$ is replaced by $\bar g_{k,m}^{\rm Hom}$ during matching and power optimization; the final performance evaluation still uses the true heterogeneous effective gain $\bar g_{k,m}^{\rm Het}$. This isolates the contributions of the algorithmic structure and the heterogeneous environment-map information.

Let $N_a=|\mathcal A|$. In the greedy stage, the $t$-th assignment evaluates at most $(M-t)\binom{K-2t}{2}N_a$ candidate region--user--power combinations. Therefore, the complexity is
$\sum_{t=0}^{M-1}(M-t)\binom{K-2t}{2}N_a$.
When $M$ is much smaller than $K$, or when the two grow on the same order, this can be conservatively upper-bounded by
$\mathcal O(M^2K^2N_a)$.
If $I_{\rm swap}$ local swap attempts are performed, the additional complexity is $\mathcal O(I_{\rm swap}M^2N_a)$. Hence, the overall complexity upper bound is
$\mathcal O\left(M^2K^2N_a+I_{\rm swap}M^2N_a\right)$,
which is polynomial and significantly lower than exhaustive joint matching and continuous power search.

\section{Simulation Results and Discussion}

The simulation setting consists of $M=4$ EDMA regions and $K=16$ candidate users. The bandwidth is $B=1$ MHz, the path-loss exponent is $\alpha=2.2$, the residual SIC factor is $\xi=0.02$, and the strong-user power coefficient is searched over $a_s\in[0.05,0.45]$ with a step size of $0.01$. The NLoS residual power coefficient is $\chi=0.10$, and the inter-region interference scaling factor is $\eta=2$. Unless otherwise specified, the main figures use $\lambda=0.5$; in the throughput--fairness operating-region figure, $\lambda$ varies from $0$ to $0.95$ with a step size of $0.05$. All results are averaged over 500 Monte Carlo trials.

The heterogeneous environment map used in the simulations follows a corridor/shelf layout, where links passing through an open corridor experience weaker blockage, whereas links traversing obstacle-dense regions experience stronger blockage. Unless otherwise specified, the blockage factors are set to $\beta_{\rm open}=0.15$, $\beta_{\rm normal}=1$, and $\beta_{\rm blocked}=5$, and the global blockage parameter is $\phi=0.02$. For the environment-heterogeneity sensitivity analysis, $\beta_{\rm open}$ is fixed while $\beta_{\rm blocked}$ is varied, giving the heterogeneity ratio
$\kappa=\beta_{\rm blocked}/\beta_{\rm open}$.
All comparison schemes use the same user distributions, SNR settings, power-search set, and true heterogeneous effective-channel evaluation model.

\begin{figure}[!t]
    \centering
    \includegraphics[width=0.95\linewidth]{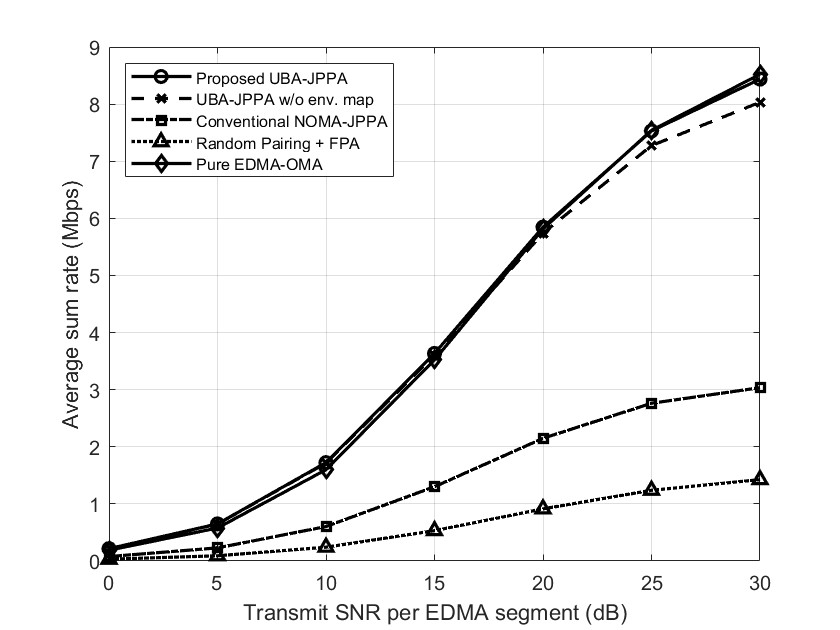}
    \caption{Average system sum rate versus SNR, with $\lambda=0.5$.}
    \label{fig:sumrate}
\end{figure}

The benchmarks include UBA-JPPA without the environment map, conventional NOMA-JPPA, random pairing with fixed power allocation, and pure EDMA-OMA. UBA-JPPA without the environment map uses the homogeneous effective gain $\bar g_{k,m}^{\rm Hom}$ during matching and power optimization, while its final performance is still evaluated using the true heterogeneous effective gain $\bar g_{k,m}^{\rm Het}$. Conventional NOMA-JPPA mainly uses homogeneous effective service-link gains for pairing and power allocation and does not explicitly exploit the heterogeneous environment map. The random-pairing baseline uses fixed power allocation with $a_s=0.2$ and $a_w=0.8$ and is evaluated under the true heterogeneous effective channel.

For the pure EDMA-OMA benchmark, the two scheduled users in each EDMA region are orthogonalized in time or frequency. The OMA rate of user $u\in\{s_m,w_m\}$ in region $m$ is computed as
\begin{equation}
    R_{u,m}^{\rm OMA}=\tau_{u,m}B\log_2\left(1+
    \frac{\rho \bar g_{u,m}}{I_{u,m}^{\rm OMA}+1}\right),
    \label{eq:omarate}
\end{equation}
where $\tau_{s,m}+\tau_{w,m}=1$ and equal resource allocation $\tau_{s,m}=\tau_{w,m}=0.5$ is adopted. The inter-region interference term $I_{u,m}^{\rm OMA}=\eta\sum_{n\neq m}\rho\bar g_{u,n}$ uses the same true heterogeneous effective-channel model as in NOMA. This OMA rate expression follows from the Shannon-type rate formula under orthogonal resource partitioning \cite{cover_thomas}.

\begin{figure}[!t]
\centering
\includegraphics[width=0.95\linewidth]{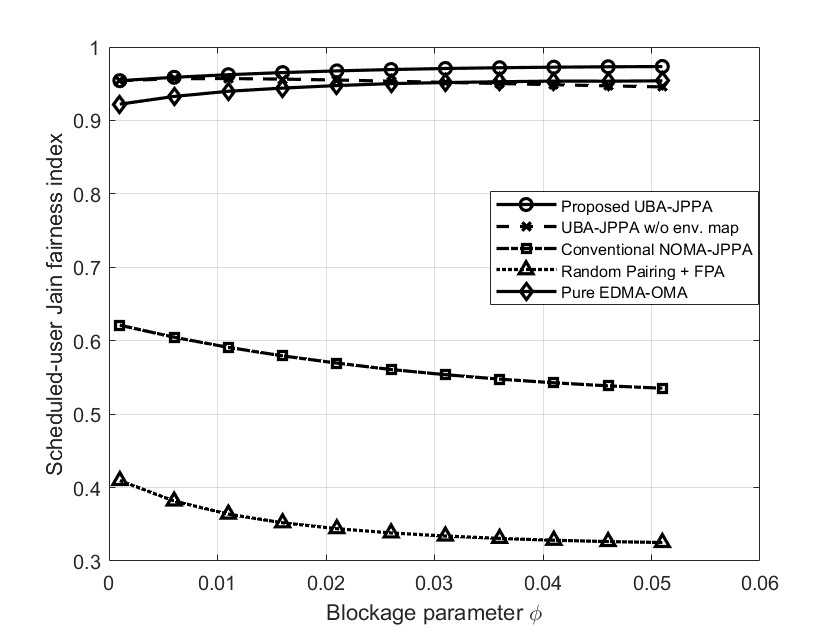}
\caption{Scheduled-user Jain fairness index versus blockage parameter $\phi$, with SNR $=20$ dB and $\lambda=0.5$.}
\label{fig:fairnessphi}
\end{figure}

\subsection{Sum-Rate Performance}

Fig. \ref{fig:sumrate} shows the average sum rate versus SNR. The sum rates of all schemes increase with SNR. The proposed UBA-JPPA consistently outperforms UBA-JPPA without the environment map, conventional NOMA-JPPA, and random pairing with fixed power allocation. This indicates that the heterogeneous environment map helps the system select user--region combinations that are more suitable for NOMA multiplexing and subject to lower inter-region leakage. Conventional NOMA-JPPA outperforms random pairing, confirming the necessity of channel-aware pairing and power optimization. Furthermore, the gain of the proposed scheme over conventional NOMA-JPPA indicates that, in EDMA systems with heterogeneous propagation environments, pairing based solely on homogeneous distance-dependent channel gains is insufficient.

At the same time, under the considered parameter settings, pure EDMA-OMA remains strongly competitive in the high-SNR region and may approach or exceed the proposed NOMA scheme. This is because EDMA-OMA avoids intra-pair NOMA interference and residual errors from imperfect SIC, although each user only occupies half of the orthogonal resource within its region. This result indicates that the goal of this paper is not to prove that EDMA-NOMA universally outperforms EDMA-OMA, but rather to verify whether the environment map can consistently improve resource allocation within the EDMA-NOMA framework.

\subsection{Fairness Performance}

Fig. \ref{fig:fairnessphi} shows the scheduled-user Jain fairness index versus the blockage parameter $\phi$. The proposed UBA-JPPA achieves the highest or near-highest fairness for all $\phi$ values and consistently outperforms the NOMA baselines without the environment map and the conventional NOMA baseline. This indicates that the environment map helps avoid user pairings that suffer from strong inter-region leakage or large intra-pair rate imbalance, thereby reducing rate disparity among scheduled NOMA users.

As $\phi$ increases, the environmental blockage effect becomes stronger, and the effective-gain disparity among different links becomes more pronounced. Both the proposed scheme and the baseline without the environment map maintain high fairness, but the proposed scheme still provides a consistent advantage. Pure EDMA-OMA also achieves high fairness because orthogonal transmission avoids intra-pair NOMA interference and residual SIC errors. In contrast, random pairing with fixed power allocation has lower fairness because it uses neither environment-aware matching nor adaptive power allocation.

\subsection{Throughput--Fairness Operating Region}

\begin{figure}[!t]
    \centering
    \includegraphics[width=1.1\linewidth]{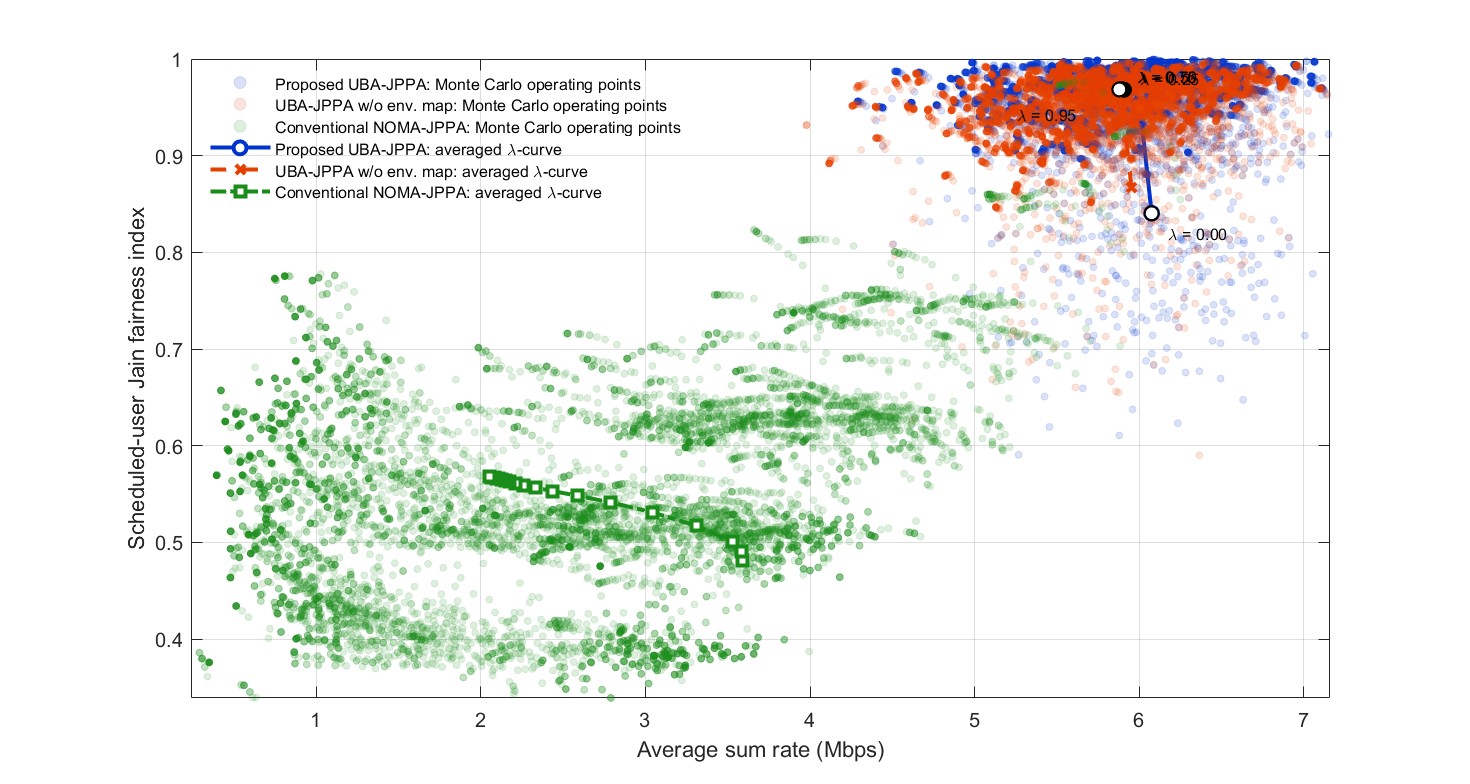}
    \caption{Throughput--fairness operating region. Translucent dots denote Monte Carlo operating points; marked curves denote averaged operating points under different $\lambda$ values; representative $\lambda$ labels indicate the trade-off direction.}
    \label{fig:tradeoff}
\end{figure}

Fig. \ref{fig:tradeoff} shows the throughput--fairness operating region. The translucent dots denote operating points from Monte Carlo realizations, while the marked curves represent averaged operating points under different $\lambda$ values. It should be noted that this figure is intended to illustrate the operating region and trade-off direction under different utility weights rather than a strict Pareto frontier. Points closer to the upper-right corner indicate a better trade-off between throughput and scheduled-user fairness. The proposed UBA-JPPA is generally located above and to the right of both UBA-JPPA without the environment map and conventional NOMA-JPPA. Since the proposed scheme and the no-environment-map scheme share the same basic algorithmic structure, this upward and rightward shift can be attributed to the additional matching information provided by the heterogeneous environment map.

\subsection{Environment-Map Ablation Study}

\begin{figure}[!t]
    \centering
    \includegraphics[width=1.05\linewidth]{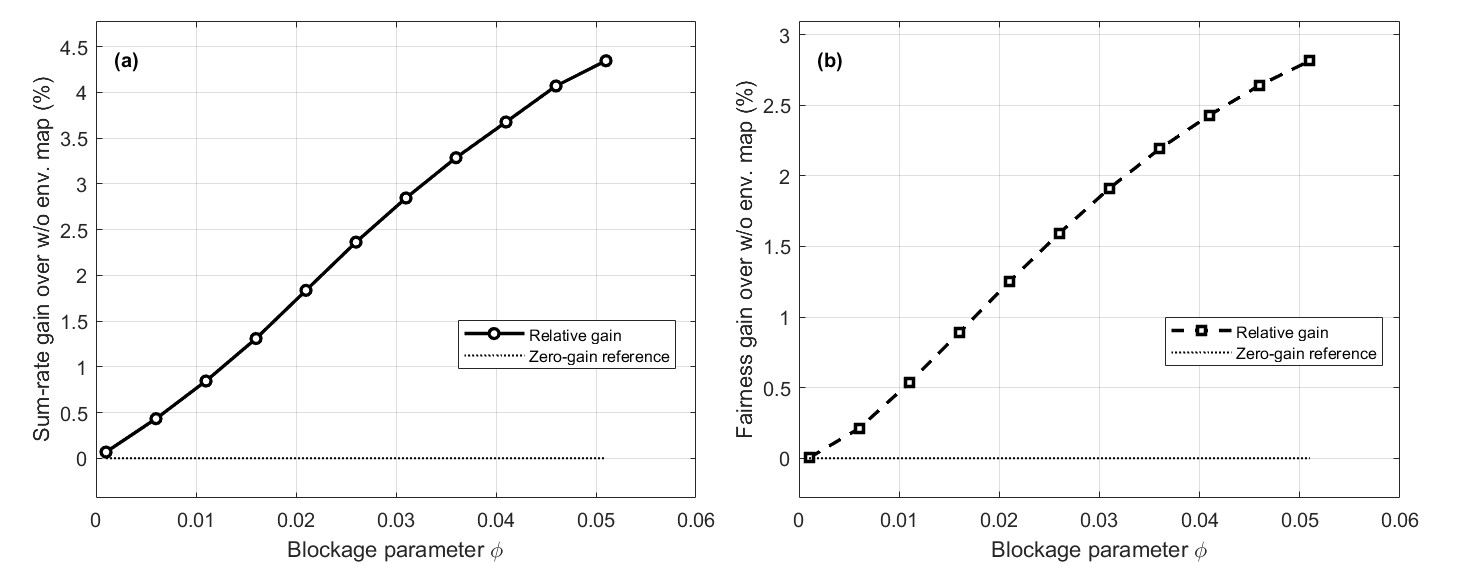}
    \caption{Environment-map ablation study: sum-rate gain and scheduled-user fairness gain of the proposed UBA-JPPA over the scheme without the environment map, with SNR $=20$ dB and $\lambda=0.5$.}
    \label{fig:ablation}
\end{figure}

Fig. \ref{fig:ablation} presents the relative gains of the proposed scheme over UBA-JPPA without the environment map. The two schemes adopt the same UBA-JPPA algorithmic structure, and their only difference lies in whether the heterogeneous environment map is used during matching and power optimization. The proposed scheme achieves positive gains in both sum rate and scheduled-user Jain fairness. Therefore, the performance improvement comes not only from ordinary NOMA pairing or power search, but also from the explicit use of heterogeneous EDMA environment-map information.

As $\phi$ increases, heterogeneous LoS/NLoS propagation differences are amplified, and the additional information provided by the environment map becomes more valuable than the homogeneous distance-based model. Therefore, within the considered parameter range, both the sum-rate gain and the fairness gain increase overall. It is worth noting that the gain magnitude is consistent yet moderate, which is in line with the more conservative average LoS/NLoS effective channel model adopted in this paper.

\subsection{Environment-Heterogeneity Sensitivity}

\begin{figure}[!t]
    \centering
    \includegraphics[width=1.05\linewidth]{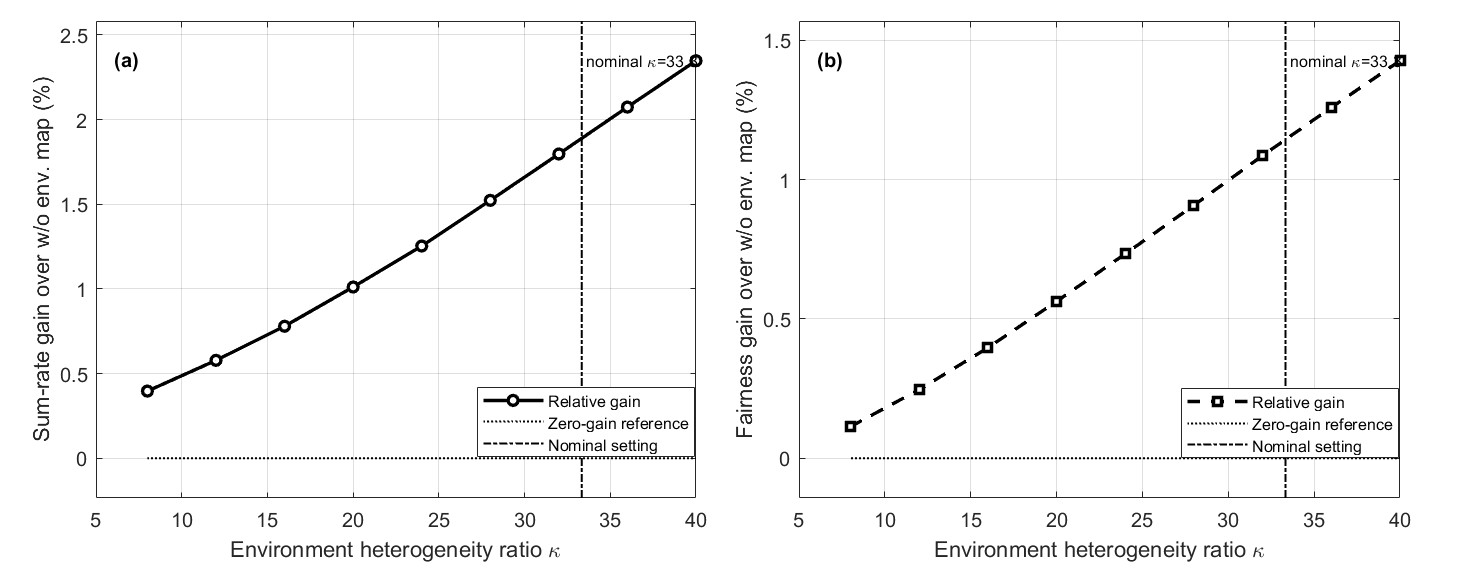}
    \caption{Environment-heterogeneity sensitivity analysis: impact of the heterogeneity ratio $\kappa$ on the environment-map gain, with SNR $=20$ dB, $\phi=0.02$, and $\lambda=0.5$.}
    \label{fig:heterogeneity}
\end{figure}

Fig. \ref{fig:heterogeneity} shows the impact of the heterogeneity ratio $\kappa$ on the environment-map gain. When SNR is 20 dB and $\phi=0.02$, the sum-rate and fairness gains of the proposed scheme over the no-environment-map baseline increase continuously over the considered range as $\kappa$ increases. This indicates that the more pronounced the difference between open and blocked regions is, the more valuable the environment map becomes. In other words, when the propagation environment is close to homogeneous, a distance-based model can already capture the main link disparities; as environmental heterogeneity increases, the explicit environment map provides link-availability information that cannot be represented by the distance model, thereby leading to more evident resource-allocation gains.

\subsection{Applicability and Limitations}

This paper adopts a static environment map and large-scale average propagation modeling. Therefore, the proposed method is more suitable for indoor, semi-indoor, or relatively stable local hotspot scenarios, such as corridors, warehouses, shelves, and partitioned areas. In scenarios with high user mobility, rapidly changing blockage, or dominant small-scale fast fading, the long-term statistical validity of the environment map may decrease. In addition, this paper assumes that the pinching-antenna radiation positions and EDMA regions have been preconfigured, and it does not jointly optimize pinching-antenna activation positions, waveguide propagation loss, or antenna radiation patterns. These factors do not change the basic conclusion regarding environment-map-assisted resource allocation, but they may affect the absolute performance of practical systems. They are therefore suitable directions for future work on robust resource allocation and joint antenna--user optimization.

\section{Conclusion}

This paper investigated heterogeneous environment-map-aware user matching and power allocation for pinching-antenna-assisted EDMA-NOMA systems. By incorporating local blockage factors into an exponential LoS probability model and further constructing average LoS/NLoS effective large-scale channel gains, the paper established the connection among environment maps, effective service links, inter-region interference, NOMA decoding rates, and resource-allocation decisions. The proposed UBA-JPPA algorithm jointly accounts for system sum rate, scheduled-user Jain fairness, SIC-favorable user pairing, and power allocation, and obtains an approximate solution through low-complexity greedy matching and local swap search. Simulation results show that the proposed scheme outperforms the EDMA-NOMA baseline without the environment map and the conventional NOMA baseline in terms of sum rate, fairness, and the throughput--fairness operating region. Ablation experiments and environment-heterogeneity sensitivity analysis further verify that the performance improvement originates from the explicit use of heterogeneous environment-map information. At the same time, the strong EDMA-OMA benchmark remains competitive in some high-SNR scenarios, indicating that the relative merits of different multiple-access modes still depend on system parameters and SIC conditions. Future work will consider robust resource allocation under map-estimation errors, small-scale fading, waveguide propagation loss, dynamic blockage, and joint pinching-antenna position optimization.

\end{document}